# Magnetic-field-assisted electron confinement and valley splitting in strained graphene

Jun-Fang Liu[1,2,§], Ke-Ke Bai[1,2,§], Jia-Bin Qiao[1,2,§], Yu Zhou[3], Jia-Cai Nie[1], Hailin Peng[3], Zhongfan Liu[2], and Lin He[1,2,*]

[1] Department of Physics, Beijing Normal University, Beijing, 100875, People's Republic of China
[2] The Center of Quantum Studies, Beijing Normal University, Beijing, 100875, People's Republic of China
[3] Center for Nanochemistry (CNC), College of Chemistry and Molecular Engineering, Peking University, Beijing, 100871, People's Republic of China

[§]These authors contributed equally to this work.
* Email: helin@bnu.edu.cn

**Spatially varying strained graphene can acquire interesting electronic properties because of the strain-induced valley-dependent gauge (pseudomagnetic) fields[1,2]. Here we report the realization of strained graphene regions located close to the step edges of Cu(111), obtained by using thermal strain engineering[3,4]. We study these strained structures with sub-nanometre-resolved scanning tunnelling microscopy and spectroscopy and identify their spatially modulated Dirac points, demonstrating the effect of overlap of Cu and graphene wave functions on the charge transfer between them[5]. By applying a magnetic field of 8 Tesla, electron confinement, as revealed by regularly spaced sharp resonances[6,7], is observed in the strained graphene. In some regions of the strained graphene, repetitive pairs of resonance peaks appear in the tunnelling spectra. This provides direct and compelling evidence for lifting of valley degeneracy due to the coexistence of both the magnetic field and the pseudomagnetic field.**

The atomically thin structure of graphene allows the existence of a variety of mechanisms which can tune its electronic properties efficiently[2,8]. A spatially varying lattice distortion in graphene, for instance, can affect the Dirac fermions like an effective magnetic field and generate partially flat bands in the band structure of graphene[1,9,10]. Because of interactions between graphene and the supporting substrates, stacking graphene on top of other materials could also change the electronic structures of this unique one-atom-thick membrane dramatically and lead to novel electronic properties not found in graphene[7,8,11-14]. Here we show how the overlap of Cu and graphene wave functions affects the position of the Dirac point in strained graphene regions located close to the step edges of Cu(111). A strong magnetic field results in electron confinement and valley splitting in the strained graphene regions, as revealed by repetitive pairs of peaks in its density of states (DOS).

Graphene films were grown on copper foil at 1000 $^o$C via a low pressure chemical vapor deposition method (see Supplementary Information A.1 for details)[4]. The Cu foils expose several different crystal facets, such as the (110) surface and the (111) surface[15]. During the cooling process of the as-grown sample, mismatch of thermal expansion coefficients between graphene and the Cu foil (the Cu substrate contracts and the graphene expands) generates an in-plane compressed stress on graphene (see Supplementary Information A.2 for details). For graphene grown on the (110) surface, the anisotropic surface stress of the Cu substrate generates one-dimensional periodic wrinkles[15] or ripples[4] in graphene. For graphene grown on the (111) surface, it exhibits quite different features. Figure 1a shows a representative, large-scale scanning tunnelling microscopy (STM) image of a graphene monolayer on the Cu(111) surface (see Supplementary Information A.3 for method of STM measurements). The in-plane compressed stress on graphene leads to the formation of strained graphene areas located close to the step edges of Cu(111), as shown in Fig. 1a-1c.

In the strained regions, the separation between graphene and Cu(111) surface $d_{ex}$ is slightly larger than that in the flat region of the terrace, as demonstrated in Fig. 2a and Fig. 2b. For convenience, we define $d_{ex} = 0$ Å in the flat region of the terrace. According to the STM measurement, an interface with the zigzag edge between the flat region and the strained region can be sharply defined down to the atomic scale (Fig. 2b). Such a structure with various graphene-metal separations provides an ideal platform to study the interaction and charge transfer between graphene and the supporting metal[5]. Figure 2c shows ten typical scanning tunnelling spectroscopy (STS) spectra recorded at different positions from the flat area to the strained graphene region on the terrace. The tunnelling spectrum gives direct access to the local DOS of the surface at the position of the STM tip. A local minimum of the tunnelling conductance, as pointed out by the arrow, is attributed to the Dirac point of the graphene. Obviously, the local Dirac point shifts from negative to positive energies when going from the flat area to the strained region, indicating effect of the graphene-Cu interaction on the charge transfer between them[5].

Figure 2d summarizes the Fermi level shifts relative to the Dirac point, $\Delta E_F$, as a function of $d_{ex}$ obtained in our experiment. For large graphene-Cu separations (in the strained graphene regions) where the graphene-Cu interaction is so weak, $\Delta E_F$ is only determined by the difference of the work function of a graphene-covered Cu ($W_{Cu}$ ~ 4.8 eV) and the work function of the free-standing graphene ($W_G$ = 4.5 eV), $i.e.$, $\Delta E_F$ = $W_{Cu}$ – $W_G$ ≈ 0.3 eV. With decreasing the graphene-Cu separations, a direct interaction between Cu and graphene, which results from the wave functions overlap between them, affects the value of $\Delta E_F$ significantly[5]. The theoretical values of $\Delta E_F$, with considering the direct interaction between Cu and graphene, are also plotted in Fig. 2d as a function of the graphene-Cu separations $d_{th}$ (see Supplementary Information A.4 for details of calculation[5]). According to the result in Fig. 2d, the Dirac point of graphene is shifted about 600 meV by the interaction between Cu and graphene. Additionally, we can conclude that the graphene-Cu separation in the flat region of Cu(111) terrace corresponds to $d_{th}$ = 3.2 ± 0.1 Å, which agrees well with its equilibrium separation, $d_{eq}$ = 3.3 Å, obtained by first-principles calculations[5].

By applying a magnetic field, the tunnelling conductance in both the flat region and the strained graphene region of Cu(111) terraces decreases, as shown in Fig. 3a and Fig. 3b. This may arise from the electronic localization in the presence of magnetic fields. The absence of Landau quantization in the flat graphene region in high magnetic fields may be attributed to the direct graphene-Cu interaction[5], which can affect the two-dimensional features of graphene's electronic properties (see Supplementary Information Figure S1 for more experimental data). In the magnetic field of 8 T, a series of discrete tunnelling peaks appear in the STS spectra recorded in the strained graphene region. The peak separation recorded at different positions ranges from about 160 meV to about 190 meV. At a fixed position, the energy spacing $\Delta E$ of the peaks is almost regularly spaced with its averaged value as about 175 meV. Such a feature precludes Landau levels in graphene systems[16-19] as the origin of these peaks (for Landau quantization in graphene monolayer, we should observe non-equally-spaced energy-level depending on the square-root of level index), instead, it reminds us possible confined electronic states in graphene[6,7,20,21]. Figure 3c shows tunnelling spectra in the magnetic field of 8 T obtained at different positions pictured in Fig. 2b. The spatially resolved STS spectra indicate that the regularly spaced resonances could only be observed in the strained graphene region, *i.e.*, the electrons are confined in the strained region. The tunnelling spectra recorded in the magnetic fields of 4 T and 6 T do not exhibit any resonances in neither the flat region nor the strained region (see Supplementary Information Figure S2 for more experimental data). Based on above experimental results, it is reasonable to conclude that both the strained structure and the strong magnetic field play vital roles in the emergence of confined electrons observed in our experiment.

Because of Klein tunnelling, which is associated with the chirality of the wavefunctions in graphene[22], scalar potential between the flat region and the strained region, as shown in Fig. 2, is unable to confine Dirac electrons. However, the gauge fields induced by the lattice deformations[1,9,10] and the magnetic fields[23] could lead to confined electronic states in graphene. The spatially resolved STS spectra shown in Fig. 3c imply that there is a large confining potential at the interface between the flat region and the strained graphene region generated by the lattice deformation and the strong magnetic field. The average energy spacing $\Delta E \sim 175$ meV of the resonances allows us to estimate the spatial scale involved in the confinement as $l \approx v_F/\Delta E \approx$ 4-6 nm. This is comparable to the size (the average width) of the strained graphene region and is also close to the value of the magnetic length generated by the magnetic field of 8 T, $l_B = \sqrt{\hbar/eB} \approx 9$ nm.

To further understand our experimental result, we calculate the effects of a confining potential on the electronic band structures of the graphene. For simplicity, we used a zigzag graphene nanoribbon with width $L = 50a$ ($a = 1.42$ Å is the nearest carbon-carbon distance), which is comparable to the magnetic length of 8 T and the average width of the strained region, to model the strained graphene region, as shown in Fig. 4a. The confining potential is assumed to be $V(x) = V_0 \left( e^{-(x-L/2)/\lambda} + e^{-(L/2-x)/\lambda} \right)$ with $V_0$ the strength and $\lambda$ the penetration depth[24]. $V(x)$ vanishes in the bulk but increases to be very large at the zigzag edge of the nanoribbon. In our system, the strength of the potential $V_0$ should depend on the magnetic field because of the fact that no confined electronic state was observed for the magnetic fields $B \leq 6$ T. In Fig. 4b, we show the electronic band spectrum for the graphene nanoribbon in the presence of the confining potential and the magnetic field 8 T. Due to the confinement, a series of regularly spaced van Hove singularities[25-27] are expected to be observed in the DOS of the nanoribbon, which agrees quite well with our experimental result. Additionally, the energy spacing of these van Hove singularities is very close to that of the DOS peaks observed in our experiment.

Besides the regularly spaced sharp resonances, we also observed repetitive pairs of tunnelling peaks in the spectra in some regions with large lattice deformations, as shown in Fig. 3c. Such a feature in principle could be expected due to valley degeneracy in graphene (the honeycomb lattices of the two-dimensional atomic crystals result in two inequivalent Dirac cones, commonly called K and K′ valley, centered at the opposite corners of the hexagonal Brillouin zone). Figure 4c shows a representative atomic structure of such a strained region where we could observe splitting of the tunnelling peaks. For the spectra recorded within the red ellipse, each resonance splits into two peaks separated by about 20-35 meV. We interpret the pairs of peaks observed in our experiment as valley splitting due to the coexistence of the magnetic field and the pseudomagnetic field. The flux of the strained graphene structure generated by lattice deformations can be estimated by $\Phi = (\beta h^2 / l a)\, \Phi_0$[9,10]. Here $\Phi_0$ is the quantum of flux, $h$ and $l$ are the height and width, as defined in Fig. 4c, of the strained structure, respectively, $\beta$ denotes the change in the hopping amplitude between nearest-neighbor atoms and usually $2 < \beta < 3$. In the structure shown in Fig. 4c, the pseudomagnetic field $B_S$ is estimated to be of the order of 1 T. Because of opposite signs of the pseudomagnetic field in the two valleys[2], the total magnetic fields in the K and K′ valleys are $B - B_S$ and $B + B_S$, respectively. Then, it is reasonable to expect different confining potential strengths $V_0(B-B_S)$ and $V_0(B+B_S)$ in the strained graphene region for electrons in the K and K′ valleys. Figure 4d shows the electronic band spectrum for the graphene nanoribbon with a slightly difference of the $V_0$ for electrons in the two valleys. Obviously, the coexistence of the magnetic field and the pseudomagnetic field lifts the valley degeneracy and results in pairs of peaks in the tunnelling spectra.

In summary, we show that the coexistence of magnetic fields and pseudomagnetic fields leads to confined electronic state and valley splitting in graphene. These properties are not present when either magnetic fields or pseudomagnetic fields are applied. Our result implies that exotic effects could be introduced in graphene when we combine its different properties.

**Acknowledgements**

This work was supported by the National Basic Research Program of China (Grants Nos. 2014CB920903, 2013CBA01603, 2013CB921701), the National Natural

Science Foundation of China (Grant Nos. 11422430, 11374035, 11474022, 51172029, 91121012), the program for New Century Excellent Talents in University of the Ministry of Education of China (Grant No. NCET-13-0054), Beijing Higher Education Young Elite Teacher Project (Grant No. YETP0238).

**Author contributions**

L.H. conceived and provided advice on the experiment, analysis, and theoretical calculation. J.F.L. and K.K.B. performed the STM experiments. J.B.Q., J.F.L., and K.K.B. analyzed the data and performed the theoretical calculations. Y.Z., H.L.P., and Z.F.L. synthesized the samples. L.H. wrote the paper. All authors participated in the data discussion.

**Competing financial interests:** The authors declare no competing financial interests.

**Figure Legends**

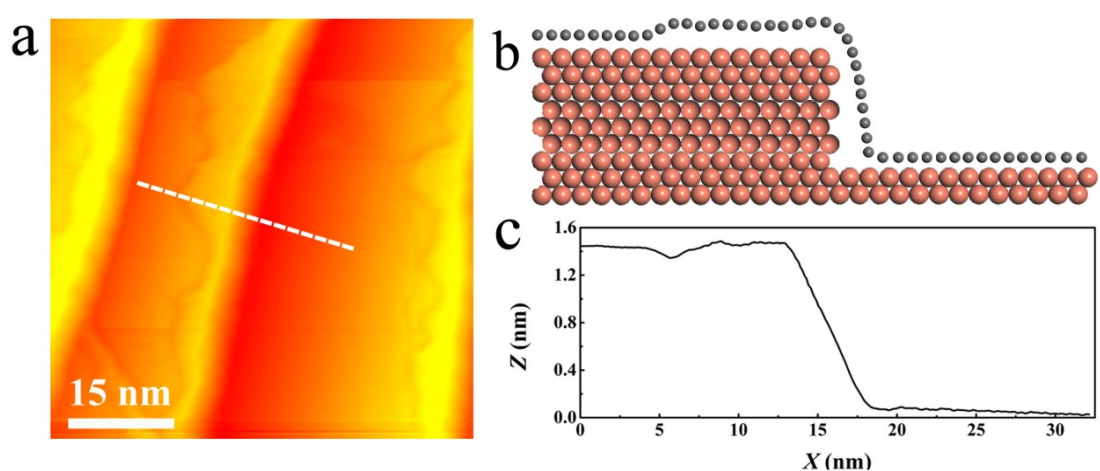

**Figure 1 | Structure of strained graphene regions on Cu(111). a,** A typical STM image of a graphene monolayer grown on Cu(111). The image has been taken at 4.5 K.

There are strained graphene areas close to the step edges of Cu(111). **b,** Schematic diagram showing a strained graphene structure located at one step of Cu(111). **c,** Height profile along the white dashed line in panel **a**.

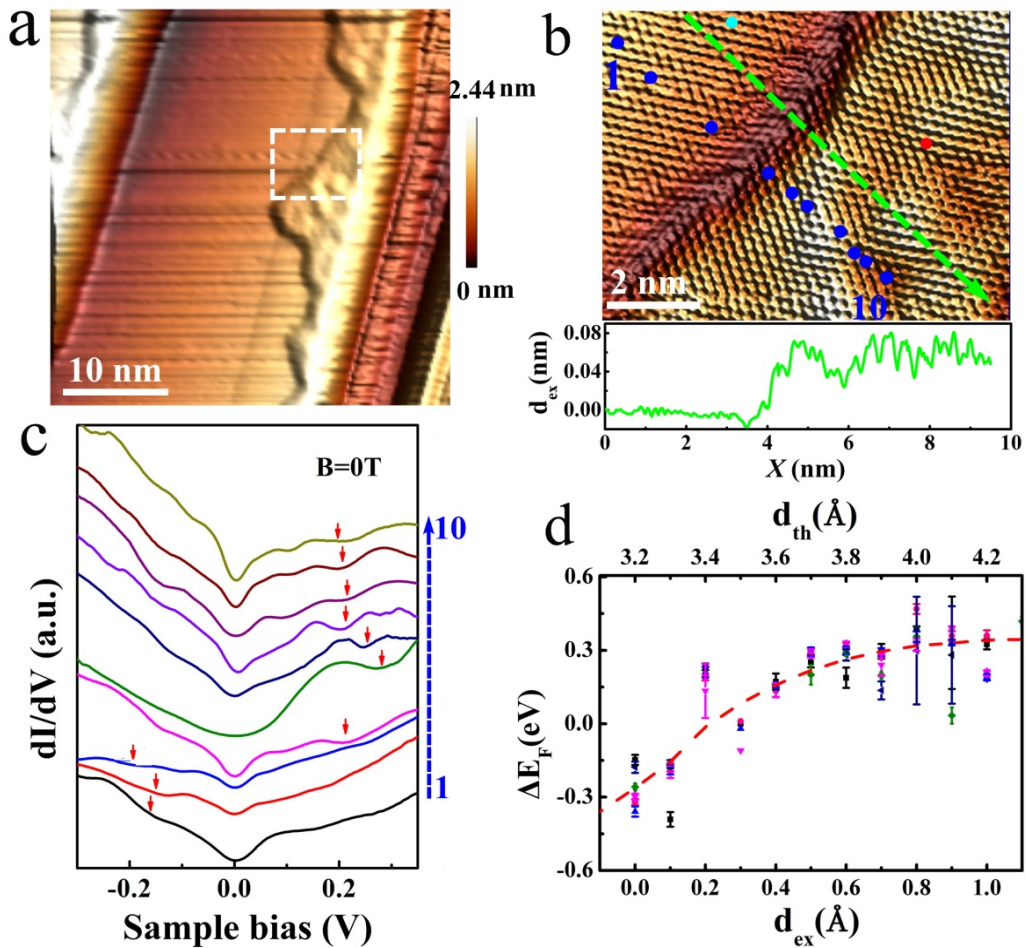

**Figure 2 | Interaction and charge transfer between graphene and Cu(111). a,** A STM topography of a strained graphene structure on a Cu(111) terrace ($V_{sample}$ = 0.38 V and $I$ = 210 pA). **b,** Upper panel: Zoom-in atomic-resolution topography obtained in the white frame in panel **a**. Lower panel: Height profile along the green dashed line in the top panel. **c,** STS spectra, *i.e.*, $dI/dV$-$V$ curves, taken at blue solid circles marked in panel **b** from 1 to 10. The red arrows denote the positions of the Dirac point at

different curves. For clarity, the curves are offset in Y-axis. a.u., arbitrary units. **d,** The solid symbols are the Fermi level shifts relative to the Dirac point, $\Delta E_F$, as a function of $d_{ex}$ (the lower $x$-axis) measured according to panel **b**. The red dashed line is the theoretical $\Delta E_F$ as a function of graphene-Cu separation $d_{th}$ (the upper $x$-axis) obtained according to Ref. 5. According to the result in panel **d**, we can conclude that the graphene-Cu separation in the flat region of Cu(111) terrace, *i.e.*, $d_{ex} = 0$ Å, corresponds to $d_{th} = 3.2 \pm 0.1$ Å.

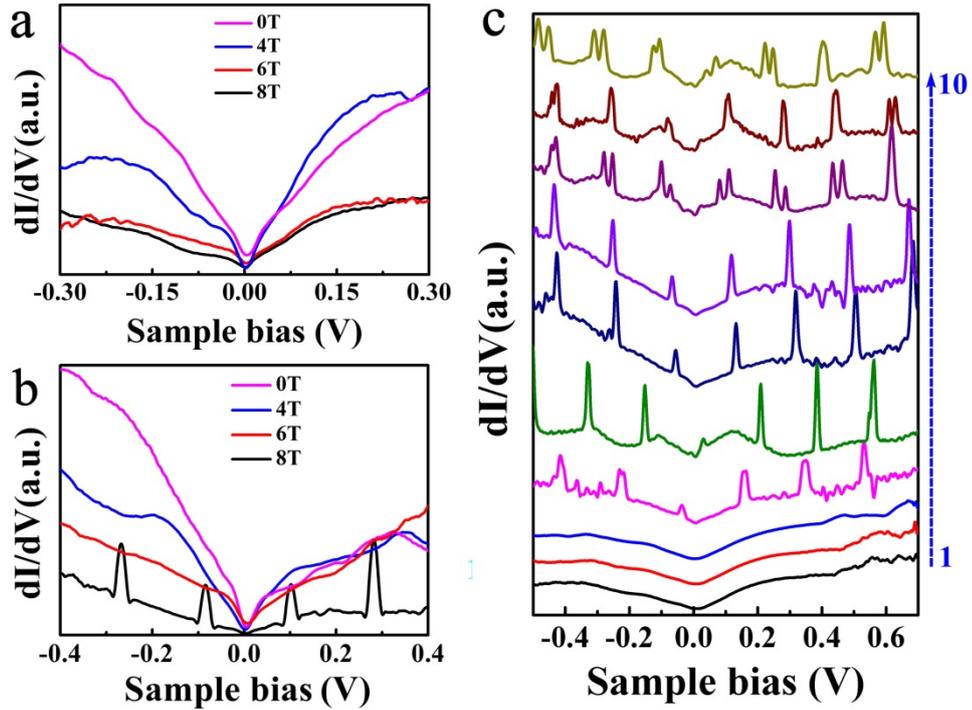

**Figure 3 | Spatially resolved STS in magnetic fields. a,** Four typical STS spectra recorded in different magnetic fields in the flat region of a Cu(111) terrace (at green solid circle marked in Figure 2b). **b,** STS spectra taken in different magnetic fields at red solid circle marked in Figure 2b. **c,** STS spectra obtained at different positions (as marked in Fig. 2b) in a magnetic field of $B = 8$ T. For clarity, the curves are offset in

Y-axis. The regularly spaced sharp resonances can only be observed in the strained graphene region. The pairs of peaks in the spectra are a sign of valley splitting.

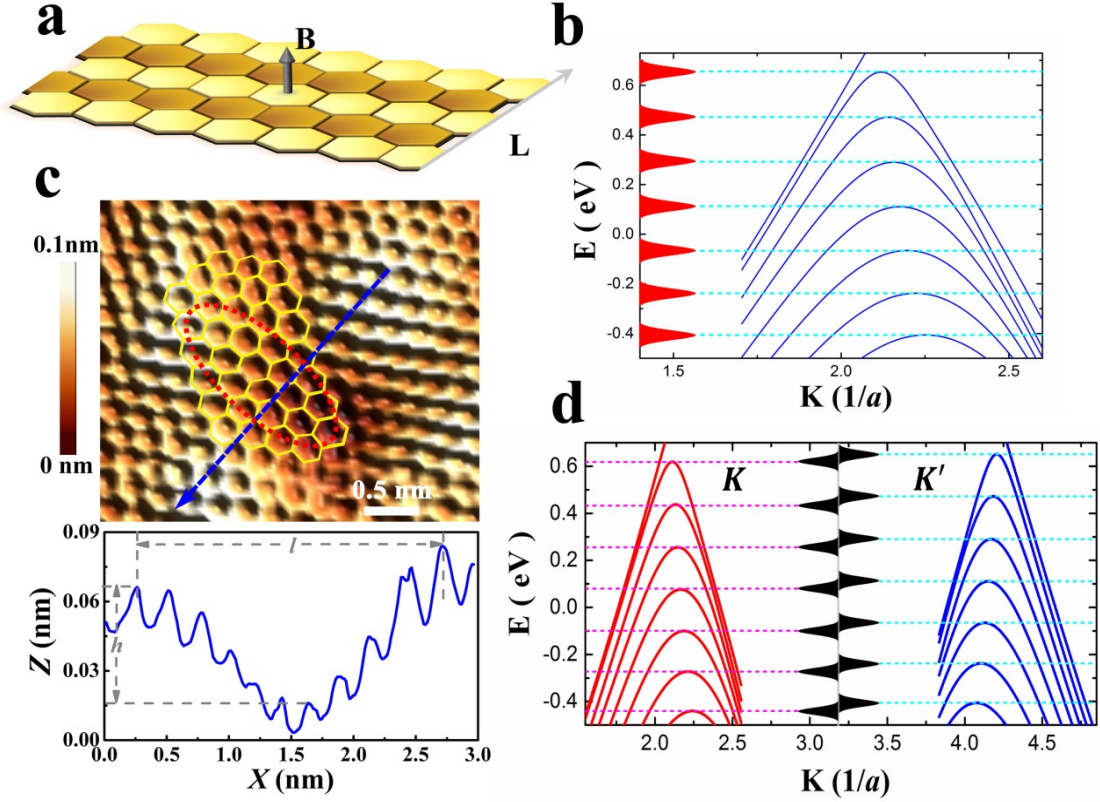

**Figure 4 | Theoretical band structures of a strained zigzag graphene nanoribbon.**
**a,** A schematic diagram for a zigzag graphene nanoribbon with width $L$. A magnetic field $B$ is applied in the perpendicular direction of the nanoribbon. **b,** Energy spectrum of a zigzag graphene nanoribbon with width $L = 50a$ and $B = 8$ T. In the calculation, a confining potential $V(x) = V_0 \left( e^{-(x-L/2)/\lambda} + e^{-(L/2-x)/\lambda} \right)$ with $V_0/t = 0.01$ and $\lambda = 9a$ is applied in the nanoribbon (here $t \approx 3$ eV is the nearest-neighbor hopping energy). The corresponding van Hove singularities in the DOS are shown in the left Y-axis. **c,** Upper panel: a zoom-in atomic-resolution STM image of the strained graphene region. The atomic structure of the graphene monolayer is overlaid onto the STM image. The red ellipse denotes a typical region where repetitive pairs of tunnelling peaks are observed in the spectra. Lower panel: the height profile along the blue arrow in the upper panel. $l$ and $h$ denote the width and the height of the strained region. **d**, Energy spectrum of the zigzag graphene nanoribbon in the presence of both the magnetic field and the pseudomagnetic field. The degeneracy of the corresponding van Hove singularities generated in the $K$ valley and the $K'$ valley are lifted.